\documentclass{article}
\usepackage[utf8]{inputenc}
\newcommand{\R}{\ensuremath{\mathbb{R}}}
\newcommand{\N}{\ensuremath{\mathbb{N}}}
\usepackage{wrapfig}
\usepackage{overpic}
\usepackage{authblk}
\usepackage{amsmath}
\usepackage{amssymb}
\usepackage[sorting = none, backend = biber, style=numeric-comp]{biblatex}
\usepackage{hyperref}
\usepackage{geometry}
\title{Parallel Quantum Annealing}

\author[1*]{Elijah Pelofske}
\author[2,**]{Georg Hahn}
\author[1,3]{Hristo N.\ Djidjev}
\affil[1]{Los Alamos National Laboratory CCS-3 Information Sciences, Los Alamos, NM 87545, USA}
\affil[2]{Harvard T.H.\ Chan School of Public Health, Boston, MA 02115, USA}
\affil[3]{Institute of Information and Communication Technologies, Bulgarian Academy of Sciences, Sofia, Bulgaria}
\affil[*]{epelofske@lanl.gov}
\affil[**]{ghahn@hsph.harvard.edu}

\date{\vspace{-5ex}}

\begin{document}
\maketitle

\begin{abstract}
    Quantum annealers of D-Wave Systems, Inc., offer an efficient way to compute high quality solutions of NP-hard problems.
    This is done by mapping a problem onto the physical qubits of the quantum chip, from which a solution is obtained after quantum annealing. However, since the connectivity of the physical qubits on the chip is limited, 
    a minor embedding of the problem structure onto the chip is required. In this process, and especially for smaller problems, many qubits will stay unused. We propose a novel method, called parallel quantum annealing, to make better use of available qubits, wherein either the same or several independent problems are solved in the same annealing cycle of a quantum annealer, assuming enough  physical qubits are available to embed more than one problem. Although the individual solution quality may be slightly decreased when solving several problems in parallel (as opposed to solving each problem separately), we demonstrate that our method may give dramatic speed-ups in terms of the Time-to-Solution (TTS) metric for solving instances of the Maximum Clique problem when compared to solving each problem sequentially on the quantum annealer. Additionally, we show that solving a single Maximum Clique problem using parallel quantum annealing reduces the TTS significantly. 
\end{abstract}
\section{Introduction}
\label{sec:intro}
Quantum annealers manufactured by D-Wave Systems, Inc. are designed to compute approximate high-quality solutions of NP-hard problems using a process called \textit{quantum annealing}. In particular, the annealers of D-Wave Systems are specialized quantum machines to minimize discrete functions that can be expressed in the form
\begin{align}
    f(x_1,\ldots,x_n) = \sum_{i=1}^n h_i x_i + \sum_{i<j} J_{ij} x_i x_j,
    \label{eq:hamiltonian}
\end{align}
where the linear weights $h_i \in \R$ and the quadratic couplers $J_{ij} \in \R$, $i,j \in \{1,\ldots,n\}$, are specified by the user and define the problem under investigation. The variables $x_i$ are unknown and take two values only: if $x_i \in \{0,1\}$ the function in eq.~\eqref{eq:hamiltonian} is called a \textit{quadratic binary optimization problem} (\textit{QUBO}), and if $x_i \in \{-1,+1\}$ it is called an \textit{Ising problem}. Both formulations are computationally equivalent.

The time evolution of any quantum system is characterized by an operator called the \textit{Hamiltonian}. For the D-Wave quantum chip, it is given by
\begin{align}
    H(s)=-\frac{A(s)}{2}\sum_{i=1}^n \sigma^x_i +\frac{B(s)}{2} \left( \sum_{i=1}^n h_i\sigma^z_i + \sum_{i\leq j} J_{ij} \sigma^z_i \sigma^z_j \right).
    \label{eq:annealing}
\end{align}
This operator is interpreted as follows. The first term encodes an equal superposition of all states (making each bit string equally likely). The second term of eq.~\eqref{eq:annealing} encodes 
the problem to be solved, given in eq.~\eqref{eq:hamiltonian}, which is fully determined through its linear and quadratic couplers $h_i$ and $J_{ij}$, respectively. During annealing, the quantum system slowly transitions from an equal superposition of all states to one whose ground state encodes the implemented problem to be solved. This is realized with the help of a so-called \textit{anneal path}, given by the functions $A(s)$ and $B(s)$. At the start of the anneal, $A(s)$ is large and $B(s)$ is small, and during annealing $A(s)$ decreases to zero while $B(s)$ increases to some maximal value. During the annealing process, it is expected that the system stays in a ground state, thus allowing to read off a low-energy (optimal or near-optimal) solution of the implemented problem upon termination.

\begin{figure*}
    \begin{overpic}[width=0.4\textwidth]{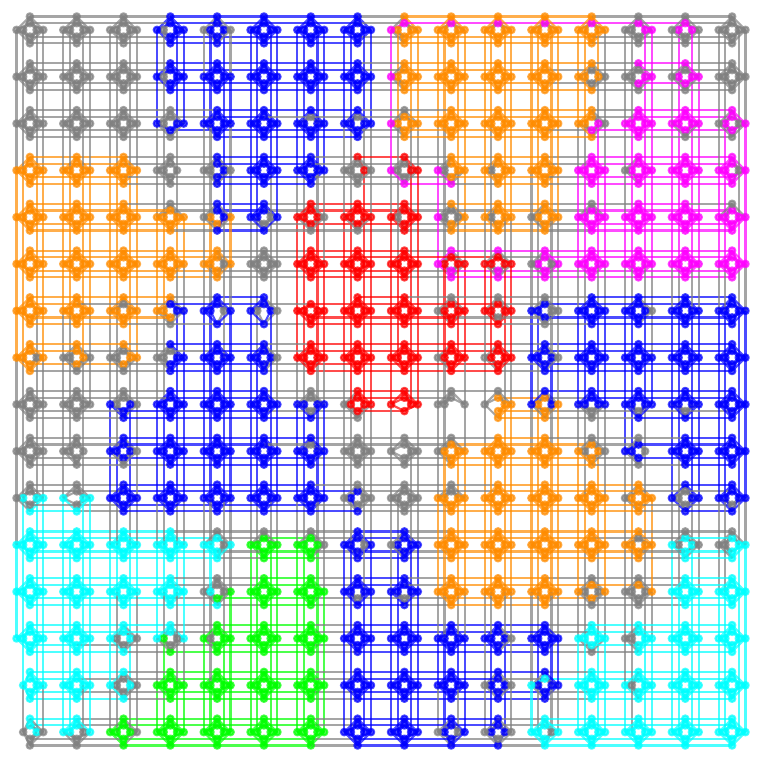}
    \put(46,-3.5){\textbf{(a)}}
    \end{overpic}\hfill
    \begin{overpic}[width=0.4\textwidth]{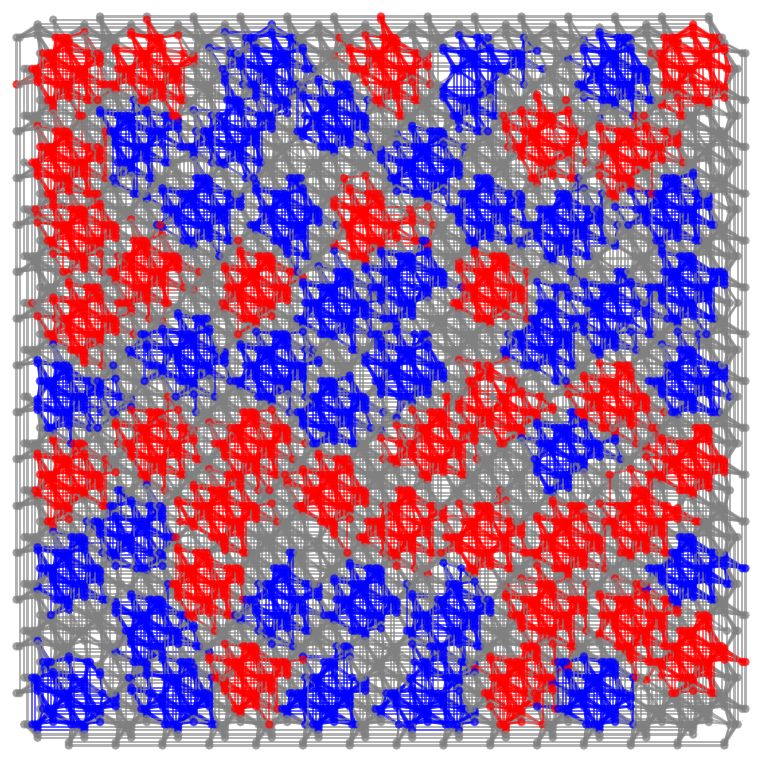}
    \put(46,-3.5){\textbf{(b)}}
    \end{overpic}
    \caption{Clique of size $20$ embedded $12$ times into the hardware of DW~2000Q, which uses Chimera graph topology (a), and $68$ times into DW~Advantage, which has a Pegasus hardware graph (b). The coloring is arbitrary and used to highlight the separation of different embeddings, with unused qubits and couplers in grey. Note that the actual embeddings do not make use of all of the physical couplers between the qubits used in the embedding; in theses figures we are showing the subgraphs induced by the qubits used in each embedding.}
    \label{fig:cliques}
\end{figure*}

To minimize eq.~\eqref{eq:hamiltonian} on the D-Wave quantum annealer, the following steps are typically required:

(i) After expressing the problem to be solved as a minimization problem of the type of eq.~\eqref{eq:hamiltonian}, one can represent eq.~\eqref{eq:hamiltonian} as a \textit{problem graph} $P$ itself having $n$ vertices, one for each variable $x_i$, $i \in \{1,\ldots,n\}$. Each vertex $i$ has a vertex weight $h_{i}$. Each non-zero term $J_{ij} x_i x_j$ becomes an edge between vertices $i$ and $j$ with edge weight $J_{ij}$.

(ii) The problem graph $P$ is mapped onto the graph of qubit connectivity of the quantum annealer. Since the D-Wave annealers offer a fixed number of physical qubits whose connectivity usually does not match the one of the problem graph $P$, a \textit{minor embedding} of $P$ onto the qubit hardware graph is typically needed. In the embedding, it is usually the case that some logical qubits are represented by a set of physical qubits on the D-Wave hardware graph, called a \textit{chain}. Chained qubits require the specification of a coupler weight as well, called the \textit{chain strength} or \textit{chain weight}, to incentivize them to take the same value during annealing. The number of chained qubits per logical qubit is called the \textit{chain length}. Embedding the problem graph $P$ onto the hardware graph results in a subgraph of the hardware graph, denoted as $P'$.

(iii) D-Wave starts the annealing process by initializing all the qubits used in the embedding of $P'$ in an equal superposition, from which the system slowly transitions to the user-specified QUBO or Ising problem while aiming to keep all qubits in the ground state. Thus a solution can be read off once the system is fully transitioned to the problem Hamiltonian.

(iv) Though they represent one logical qubit, chained hardware qubits are not guaranteed to take the same value after annealing (we call those \textit{broken chains}). To arrive at a consistent solution, we need to \textit{unembed} chains, meaning that each broken chain has to be assigned one value after annealing. There is no unique way to achieve this, though D-Wave offers several default methods for unembedding.

Since computing a minor embedding is computationally intensive, the annealer is oftentimes used with a fixed precomputed embedding of a complete graph $K_s$ of size $s$, which is the maximum size that can fit onto the qubit hardware. This has the advantage that any problem graph $P$ of $s$ vertices can be easily embedded into the annealer as it is a subgraph of $K_s$. However, using a fixed complete graph embedding also has disadvantages. First, using a fixed complete embeddding might lead to a poorer solution quality than using tailored embeddings~\cite{Barbosa2021prediction}. This is due to increased problem complexity, as well as longer chain lengths that distort the original problem. Second, if the problem to be solved on D-Wave does not require all (or the majority) of the hardware qubits, the D-Wave hardware is used suboptimally since qubits remain unused during annealing that could be employed to solve other problems. This second case is aggravated by the fact that the largest complete embedding which fits onto the hardware does not use all of the available qubits.

In this contribution, we show that multiple instances of either the same or different problems can be embedded and subsequently solved in parallel on a single quantum annealing chip, while suffering from only small decrease in individual solution quality (measured by the probability of finding a ground state solution). Specifically, we propose a method called \textit{parallel quantum annealing} for solving multiple problems simultaneously on a quantum annealer, which works as follows. Given a limit on the number of variables allowed per QUBO problem to be solved simultaneously, say $s \in \N$, we compute a maximum number of embeddings, say $k \in \N$, of complete graphs of order $s$ that can be placed on the quantum chip (see Figure~\ref{fig:cliques}). This will allow any $k$ QUBO problems with no more than $s$ variables each to be easily embedded into the QPU (quantum processing unit), since any problem graph of $s$ vertices can be embedded into a complete graph of order $s$. We can then solve the resulting composite QUBO in a single call on the quantum annealer and decompose the returned solution into solutions of the individual QUBOs. This is justified since, when adding together independent QUBOs (that is when adding together functions of the form of eq.~\eqref{eq:hamiltonian} that do not share any unknown variables), the solution of the composite QUBO will be the union of all individual ground states, and the bitstring yielding the minimum will be the concatenation of all individual bitstring solutions.

We apply our proposed parallel quantum annealing methodology to the Maximum Clique problem, an important NP-hard problem with applications in network analysis, bioinformatics, and computational chemistry~\cite{Bomze1999, Rossi2014, qtop}. For a graph $G=(V,E)$ with vertex set $V$ and edge set $E$, a \textit{clique} in $G$ is any subgraph $C$ of $G$ that is \textit{complete}, i.e., there is an edge between each pair of vertices of $C$. The Maximum Clique problem asks us to find a clique of maximum size. A formulation of the Maximum Clique problem in the form of eq.~\eqref{eq:hamiltonian} is available in the literature~\cite{qtop, Chapuis2019}. In this article, we will consider both the D-Wave 2000Q at Los Alamos National Laboratory (referred to as \textit{DW~2000Q}), and the newer D-Wave Advantage\_System~1.1 (referred to as \textit{DW~Advantage}) which we accessed through D-Wave Leap.

Multiple problems of the same type may need to be solved in various scenarios. For instance, decomposition methods~\cite{decomposition_algos_journal} for solving optimization problems such as Maximum Clique allow one to divide an input problem that is too large to fit on the QPU into many smaller problems from which a solution of the original problem can be constructed. Methods in computer vision~\cite{Li2015}, genomics~\cite{Maenhout1463}, and protein structure analysis~\cite{ChapuisBADL15} have been shown to lead to solving multiple Maximum Clique problems. Also, a hard instance of the same problem (e.g., Maximum Clique) may need to be solved multiple times on the quantum annealer in order to find an optimal or high-quality solution, and all such executions are independent and can be run in parallel with our proposed method.

The article is structured as follows. We start with a brief literature review in Section~\ref{sec:literature}. Experimental results are reported in Section~\ref{sec:experiments}, followed by a discussion in Section~\ref{sec:discussion}. In Section~\ref{sec:methods} we highlight the rationale behind solving problems in parallel on the D-Wave hardware architecture, in particular the choice of the embedding. We also provide a generalization of the TTS (Time-To-Solution) formula we employ to measure the TTS metric for parallel and sequential problem solving.

\subsection{Literature review}
\label{sec:literature}
Ising spin glass models have been extensively studied in the literature, for instance with respect to the stability of the system, its phase transitions, and its magnetization distribution~\cite{Ray1989}. To find the ground state of such spin glass models, as well as for other problem Hamiltonians, quantum annealing has been shown to find ground states with higher probability than classical (thermal) simulated annealing~\cite{Kadowaki1998}.

The fact that a quantum system will stay near its ground state if its Hamiltonian varies slowly enough led to adiabatic quantum algorithms~\cite{Farhi2001}, for which it was hypothesized early that they might be able to outperform classical algorithms on instances of NP-complete problems. In particular, quantum annealing can be shown to converge to the ground state energy at a faster rate than its classical counterpart~\cite{Santoro2002}. A comprehensive review on combinatorial optimization problems (such as the ones studied in the present article), spin glasses, and quantum annealing via adiabatic evolution is available in the literature~\cite{Das2008}.

There are very few published results yet concerning parallelism and quantum annealing, and none of them discusses the type of parallelism we propose. Some papers consider pairing quantum computing systems with modern HPC infrastructure~\cite{Humble2021}. The focus there lays on the design of macroarchitecture, microarchitecture, and programming models needed to integrate near-term quantum computers with HPC systems, and the near-term feasibility of such systems. They do not consider, however, parallelism at the level of a quantum device.

In a different context~\cite{jalowiecki2019parallel}, the authors study the problem of simulating dynamical systems on a quantum annealer. Such systems are intrinsically sequential in the time component, which makes them difficult to simulate on a quantum computer. They propose to use the \textit{parallel in time} idea from classical computing to reformulate a problem so that it can be solved as the task of finding a ground state of an Ising model. Such an Ising model is then solved on a quantum annealer. However, unlike our work, there is only one Ising problem solved at a time.

Another approach considers a framework built upon Field Programmable Gate Array chips (FPGA) in connection with probabilistic bits to simulate Gibbs samplers\cite{aadit2021massively}. The authors use their algorithm on blocks of conditionally independent nodes of the input graph, which are updated in parallel using a quantum annealer. The authors remark that technically, all blocks have to be completed before some synchronization step is carried out; however, if not all blocks are completed, the network is still able to find exact ground states. However, the parallel architecture proposed by the authors~\cite{aadit2021massively} is not quantum.

To the best of our knowledge, we are the first to propose solving multiple independent problems in parallel on a single quantum device. In a previous contribution~\cite{pelofske2021quantum}, the parallel quantum annealing methodology has been used in a specific manner as a way to solve $255$ small QUBOs (with a maximum of $4$ variables each) simultaneously on the D-Wave 2000Q at Los Alamos National Laboratory with a relatively small number of anneals. In contrast to the previous work~\cite{pelofske2021quantum}, the present article expands on the parallel quantum annealing idea by investigating larger problem sizes, considering the NP-hard Maximum Clique problem, and applying it to the new DW~Advantage generation that uses a Pegasus qubit connection topology~\cite{boothby2020nextgeneration}.
\section{Experiments}
\label{sec:experiments}
In this section, we assess the proposed parallel quantum annealing technique by solving several Maximum Clique problems in the same D-Wave call. First, after two brief notes on the parameter setting (Section~\ref{sec:tuning}) and the Time-To-Solution metric (Section~\ref{sec:tts}), we investigate in Section~\ref{sec:accuracy_parallel} the behavior of both the ground state probability and the TTS measure as a function of the problem size for various graph densities, and for both DW~2000Q and DW~Advantage. Next, in Section~\ref{sec:comparison_sequential_parallel}, we compare parallel quantum annealing and sequential quantum annealing, again with respect to both ground state probability and the TTS measure. A comparison of parallel quantum annealing with the classical FMC solver \cite{fmc} on the Maximum Clique problem is given in Section~\ref{sec:classical}. We conclude with an assessment of how our method can help to solve a single hard problem by implementing it multiple times on the hardware in Section~\ref{sec:same_problem_improved_TTS}.

\subsection{Parameter tuning}
\label{sec:tuning}
The D-Wave annealers offer a variety of tuning parameters that can be chosen by the user before each anneal. Those encompass the annealing time, the chain strength, anneal offsets, etc. We perform a grid search on the DW~2000Q in order to find reasonable values for both the chain strength and the annealing time, with the goal to minimize the QPU time while maximizing the probability of finding the ground state of each of the embedded problems.

In all experiments, we set the annealing time to $50$ microseconds, and compute the chain strength using the \textit{uniform torque compensation} method included in the D-Wave Ocean SDK~\cite{UTC} with a UTC prefactor of $0.2$. Lastly, We set the \textit{programming thermalization} time to $0$ microseconds, and the \textit{readout thermalization} time to $0$ microseconds. We use the same parameters for the experiments with DW~Advantage as we do for the ones with DW~2000Q. However, the performance of DW~Advantage might be potentially improved by additional parameter optimization.

\subsection{Time-To-Solution for an ensemble of problems}
\label{sec:tts}
Since quantum annealers are stochastic (and heuristic) solvers, they are not guaranteed to find the ground state (the global minimum) of a problem of the type of eq.~\eqref{eq:hamiltonian} in each anneal. Therefore, to have a meaningful metric of accuracy, \textit{Time-To-Solution} is used to quantify the expected time it takes to reach an optimum solution with a $99$ percent confidence~\cite{king2015benchmarking, Barbosa2021, 10.1145/3492805.3492815, doi:10.1126/sciadv.aau0823}. It is defined, in the sequential case, as
\begin{align}
    \text{TTS}_{seq} =\frac{1}{A} (T_\text{QPU} + U) \frac{\log(0.01)}{\log(1-p) },
    \label{eq:tts1}
\end{align}
where $p$ is the probability of finding the ground state within a single anneal, $T_\text{QPU}$ is the total D-Wave QPU time (specifically the \textit{qpu-access-time}), and $U$ is the total CPU process time used to compute the unembedded solutions in seconds. Note that in eq.~\eqref{eq:tts1}, we do not include the embedding time because we compute full clique embeddings, and then save and re-use those embeddings.

In the context of the present work, a generalization of the standard TTS measure is required since we solve $K$ problems of size $N$ simultaneously on the same D-Wave chip, using a total QPU processing time of $T_\text{QPU}$ and $A$ anneals.

For each problem $i \in \{1,\ldots,K\}$ we solve simultaneously, we record the proportion of correct solutions $p_i$ among the $A$ anneals. We weigh those using the formula
$$p_K = \frac{1}{K} \sum_{i=1}^K p_i,$$
where every $p_i$ must be non-zero. We generalize eq.~\eqref{eq:tts1} as
\begin{align}
    \text{TTS}_{ens}(K) = \frac{1}{A} \left( \frac{T_\text{QPU}}{K} + U \right) \frac{\log(0.01)}{\log(1-p_K)},
    \label{eq:tts2}
\end{align}
where $U$ is the total CPU unembedding time used to unembed the solutions of each of the $K$ problems. We refer to $\text{TTS}_{ens}$ as the \textit{ensemble Time-To-Solution}, since it captures the time to reach an optimal solution for each problem in a group of problems that are solved simultaneously.

\subsection{Accuracy of parallel quantum annealing}
\label{sec:accuracy_parallel}
\begin{figure*}
    \centering
    \begin{overpic}[width=0.49\textwidth]{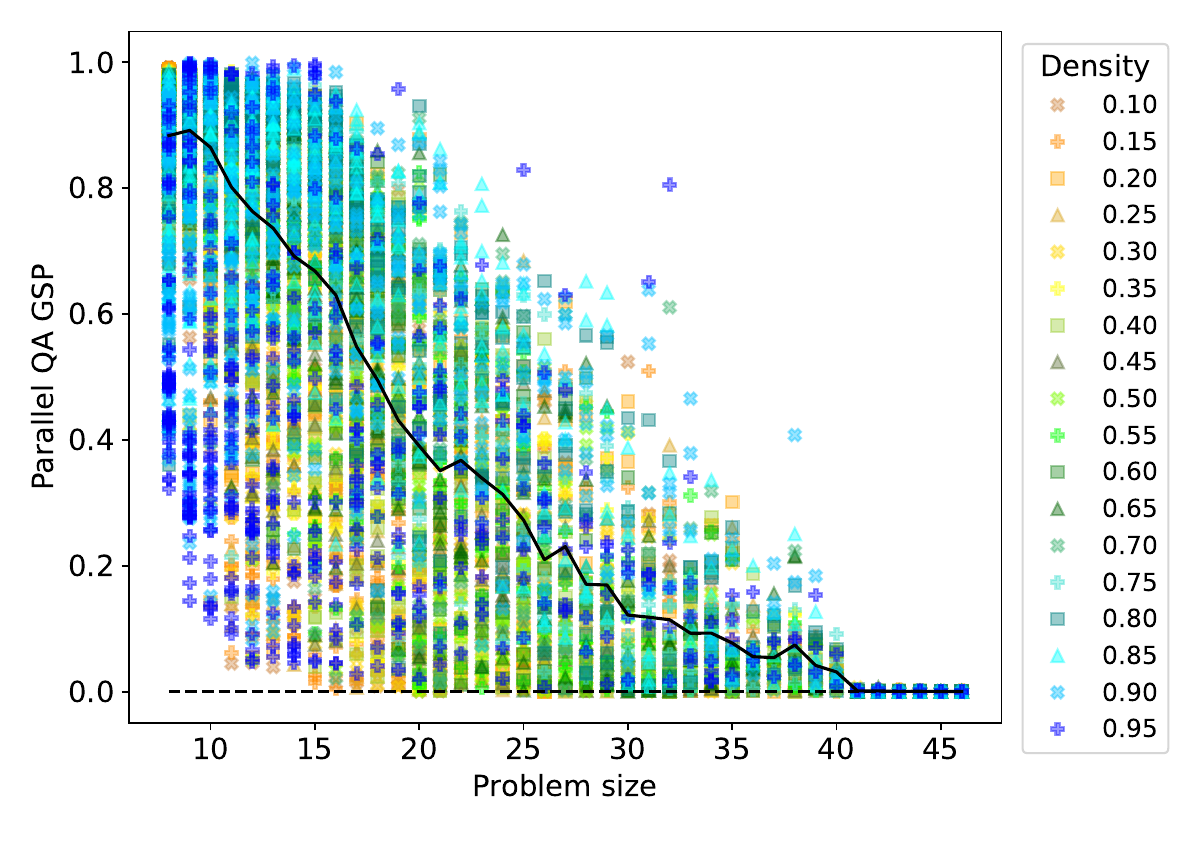}
    \put(46,-1){\textbf{(a)}}
    \end{overpic}\hfill
    \begin{overpic}[width=0.49\textwidth]{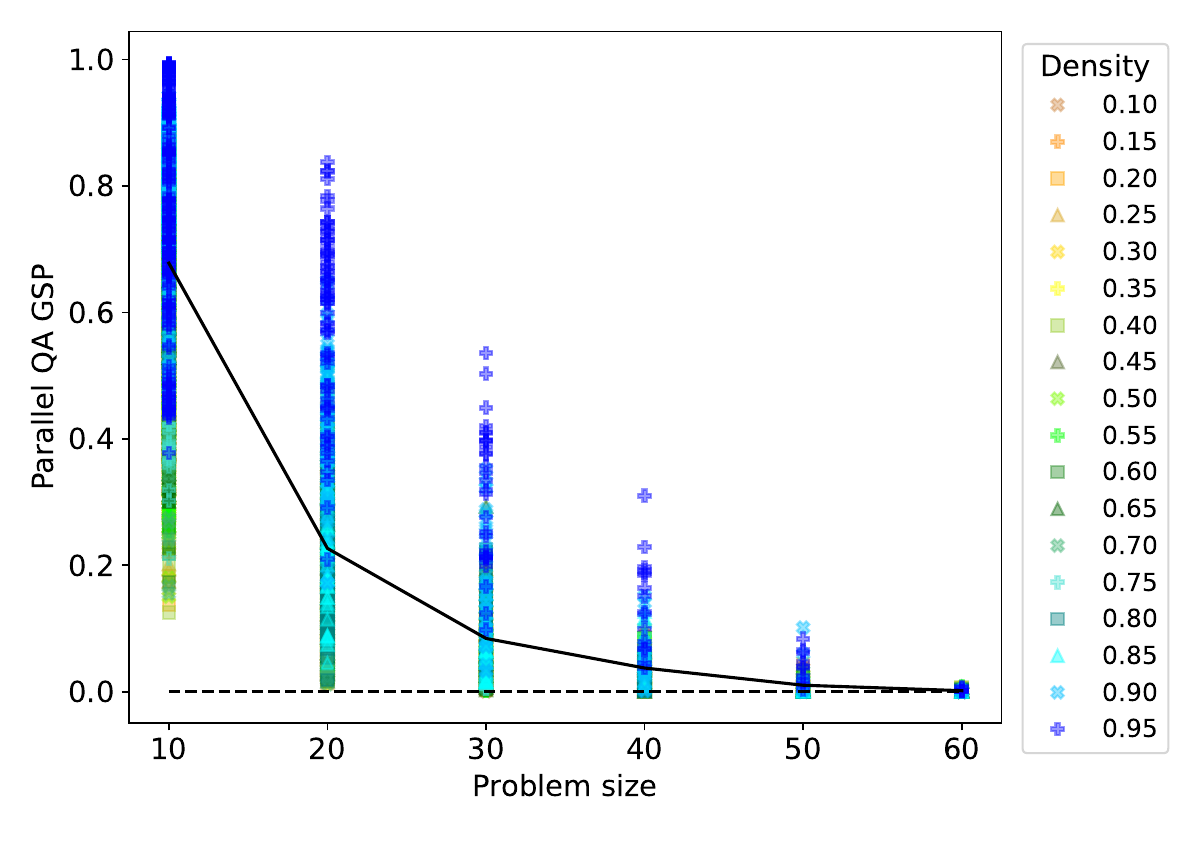}
    \put(46,-1){\textbf{(b)}}
    \end{overpic}\hfill
    \caption{Ground state probability (GSP) achieved by parallel quantum annealing for DW~2000Q (a) and for DW~Advantage (b) as a function of the problem size $N$. Graph densities range from $0.10$ to $0.95$ (color coding shown in the legend). Black solid line plots the mean parallel quantum annealing GSP as a function of problem size.}
    \label{fig:GSP_parallel_QA}
\end{figure*}

This section studies the accuracy of parallel quantum annealing per se.
Using the parameters of Section~\ref{sec:tuning}, we investigate the probability with which a ground state is found as a function of the problem size. Figure~\ref{fig:GSP_parallel_QA} shows results for both DW~2000Q (a) and for DW~Advantage (b). We observe that, as expected, DW~2000Q finds the ground state probability (GSP) more reliably for smaller problems, though the probability varies quite considerably with the graph density. Denser problem graphs are typically easier than sparser ones. For cliques of roughly size $40$ onward, DW~2000Q is unable to find the ground state. For DW~Advantage we observe a similar picture. The GSP is higher for smaller problems than for larger ones, with an even bigger spread by graph density. As before, denser problem graphs are easier to solve reliably than sparser ones. As expected, DW~Advantage is able to solve larger inputs, in particular it finds the ground state with appreciable probability of up to $40\%$ even for inputs of size $40$. 

Something we noted in these results is that the parallel quantum annealing method never found all ground state solutions in a single anneal. Instead, different sets of ground state solutions were found with each anneal - and over a sequence of anneals we eventually find the solutions to all of the problems. Interestingly, the rate at which the ground state solutions were found was not the same across all problems - indeed it was heavily biased. This bias is likely due to differences in the embeddings. 

\begin{figure}
    \centering
    \begin{overpic}[width=0.49\textwidth]{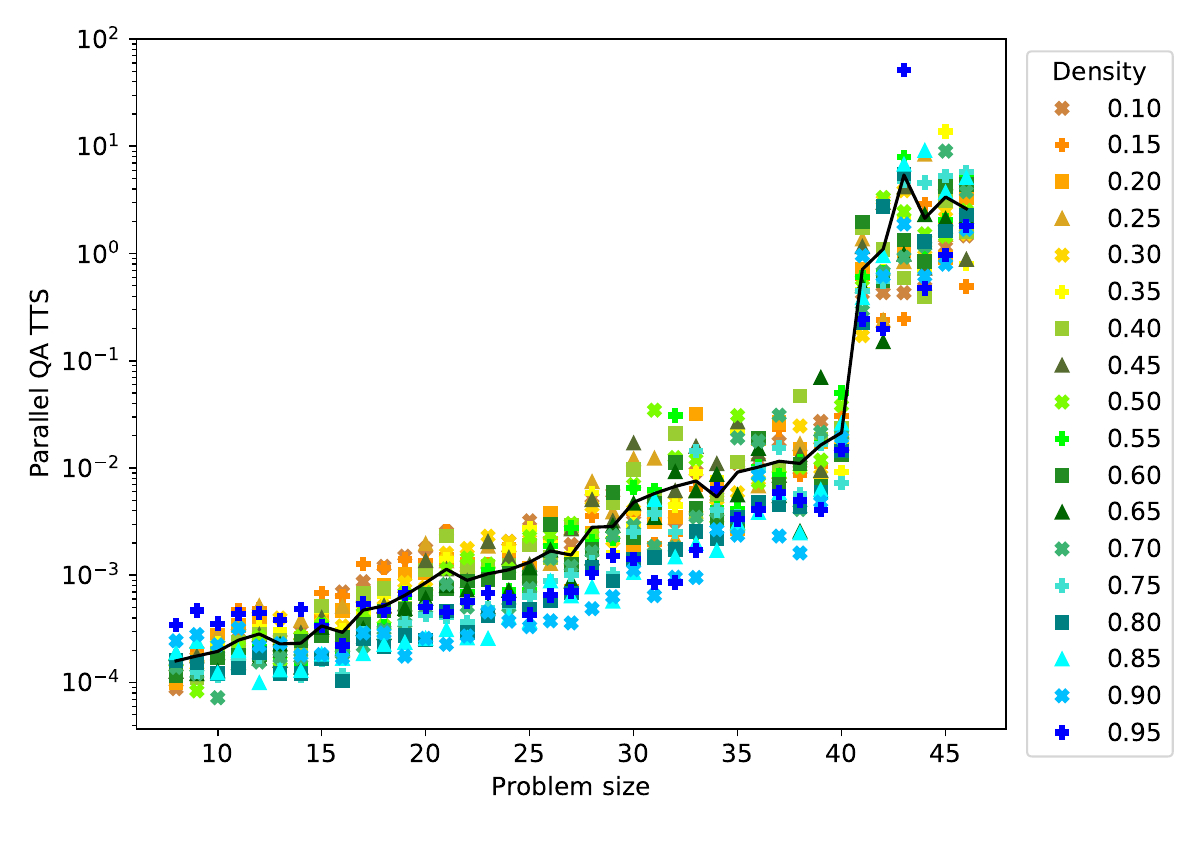}
    \put(46,-1){\textbf{(a)}}
    \end{overpic}\hfill
    \begin{overpic}[width=0.49\textwidth]{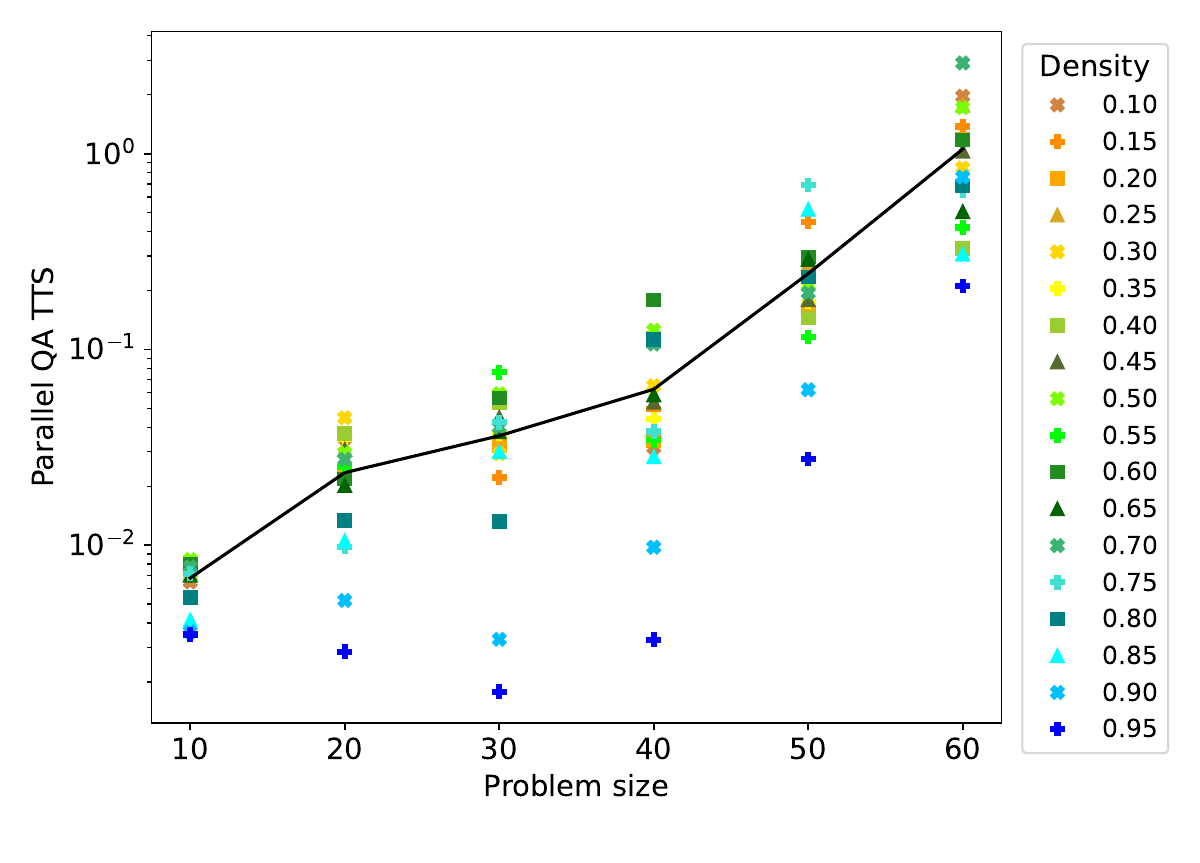}
    \put(46,-1){\textbf{(b)}}
    \end{overpic}
    \caption{Ensemble TTS for parallel quantum annealing as a function of the problem size for DW~2000Q (a) and DW~Advantage (b). Graph densities range from $0.10$ to $0.95$ (color coding shown in the legend). Black solid line plots the mean parallel quantum annealing ensemble TTS as a function of problem size.}
    \label{fig:TTS_parallel_QA}
\end{figure}

Similarly to Figure~\ref{fig:GSP_parallel_QA}, Figure~\ref{fig:TTS_parallel_QA} shows the ensemble TTS measure for parallel quantum annealing as a function of the problem size, and for various graph densities. For DW~2000Q (a), we observe that this TTS measure increases as the problem size increases, which is as expected. The reason for the "jump" at around problem size $40$ is unknown, but the behavior of D-Wave could be related to an internal hardware issue (during the time frame in which some of this data was taken on DW~2000Q, around problem sizes 38-46, there was an initially undetected temperature increase from the typical $0.015$ Kelvin of the hardware, likely leading to a poorer solution quality). In accordance with Figure~\ref{fig:GSP_parallel_QA}, dense problem inputs seem to be easier for D-Wave, and thus incur a lower TTS value. For DW~Advantage (b), we observe a similar picture. Interestingly, for almost all problem sizes that can be solved on DW~2000Q (roughly up to size $40$), the DW~2000Q is actually the better solver with respect to the TTS metric. For sizes above 40, DW~Advantage is the better solver with respect to TTS.

\subsection{Comparison of sequential to parallel quantum annealing}
\label{sec:comparison_sequential_parallel}
\begin{figure}[t]
    \centering
    \begin{overpic}[width=0.49\textwidth]{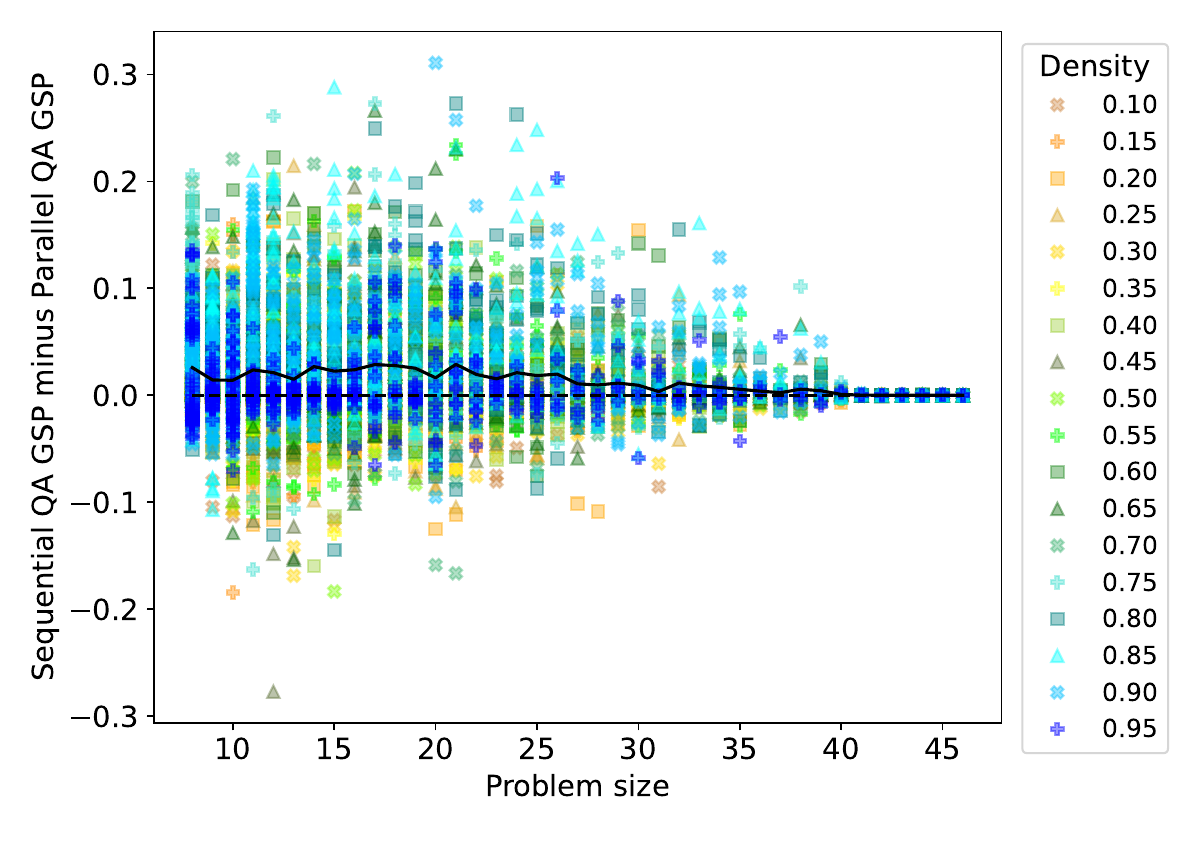}
    \put(46,-1){\textbf{(a)}}
    \end{overpic}\hfill
    \begin{overpic}[width=0.49\textwidth]{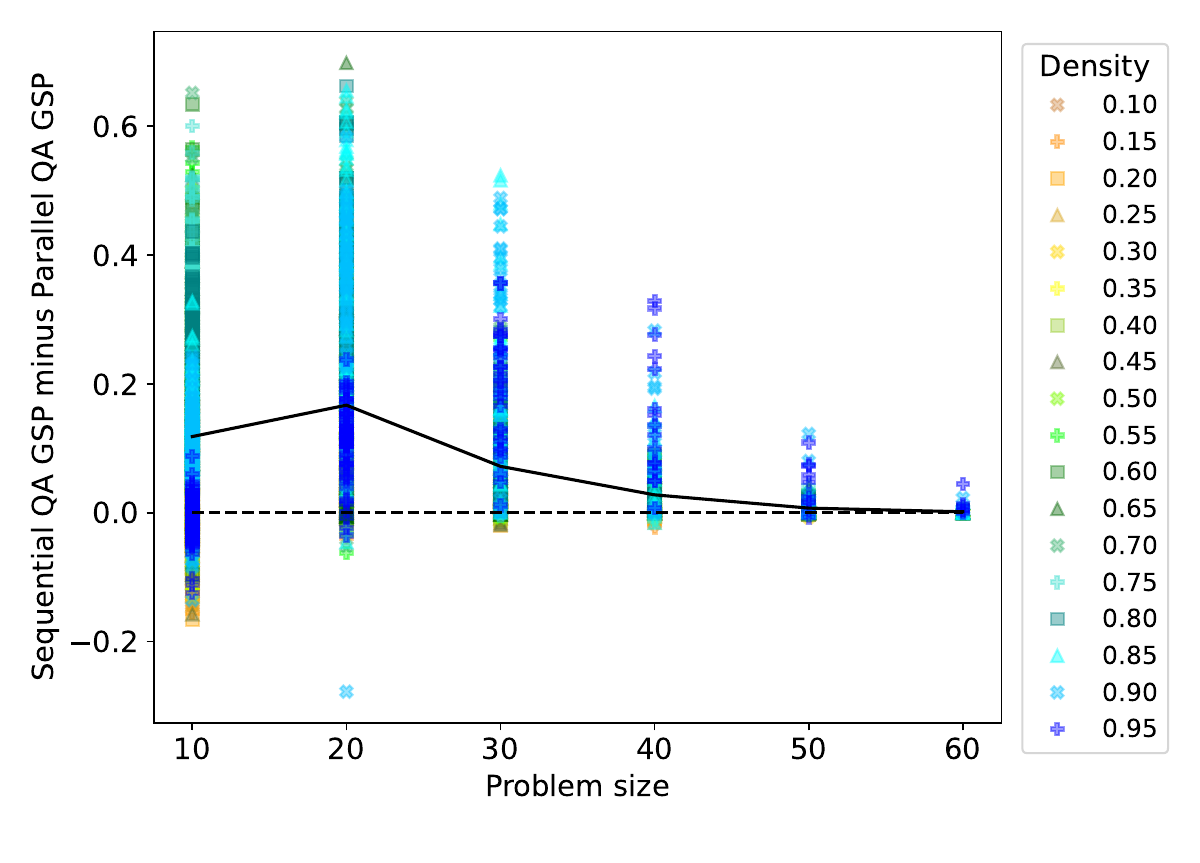}
    \put(46,-1){\textbf{(b)}}
    \end{overpic}
    \caption{Difference between the ground state probabilities (GSP) of sequential and parallel quantum annealing as a function of the problem size for DW~2000Q (a) and DW~Advantage (b). Values above zero indicate that sequential quantum annealing finds higher ground state probabilities than parallel quantum annealing. Averages indicated with a solid black line. Graph densities range from $0.10$ to $0.95$ (color coding shown in legend).}
    \label{fig:GSP_comparison}
\end{figure}
To compare the benefits and trade-offs of solving several problems in parallel on the D-Wave quantum annealers, we first investigate the difference between the GSP found by both sequential and parallel quantum annealing. Here, sequential quantum annealing refers to solving the problems separately one after the other on the D-Wave architecture. The GSP in the parallel case is simply defined per problem as the probability of finding its individual ground state. Figure~\ref{fig:GSP_comparison} shows results for DW~2000Q (a) and for DW~Advantage (b). We observe that, on average across the graph densities considered, solving problems simultaneously causes D-Wave to find a ground state slightly less frequently than when solving them sequentially. For DW~Advantage, this phenomenon is even more pronounced.

The reduced GSP we observe for parallel quantum annealing can be explained as follows. First, the problem complexity (i.e., the number of variables and quadratic terms) increases when solving multiple problems as opposed to one, potentially causing difficulties for the D-Wave annealer. Second, multiple embeddings on the chip mean that there is a closer physical proximity between the qubits used per embedding, potentially causing increased leakage and interaction between the involved qubits. However, unlike TTS, the GSP metric does not account for the fact that we are finding solutions to multiple problems.

\begin{figure}
    \centering
    \begin{overpic}[width=0.49\textwidth]{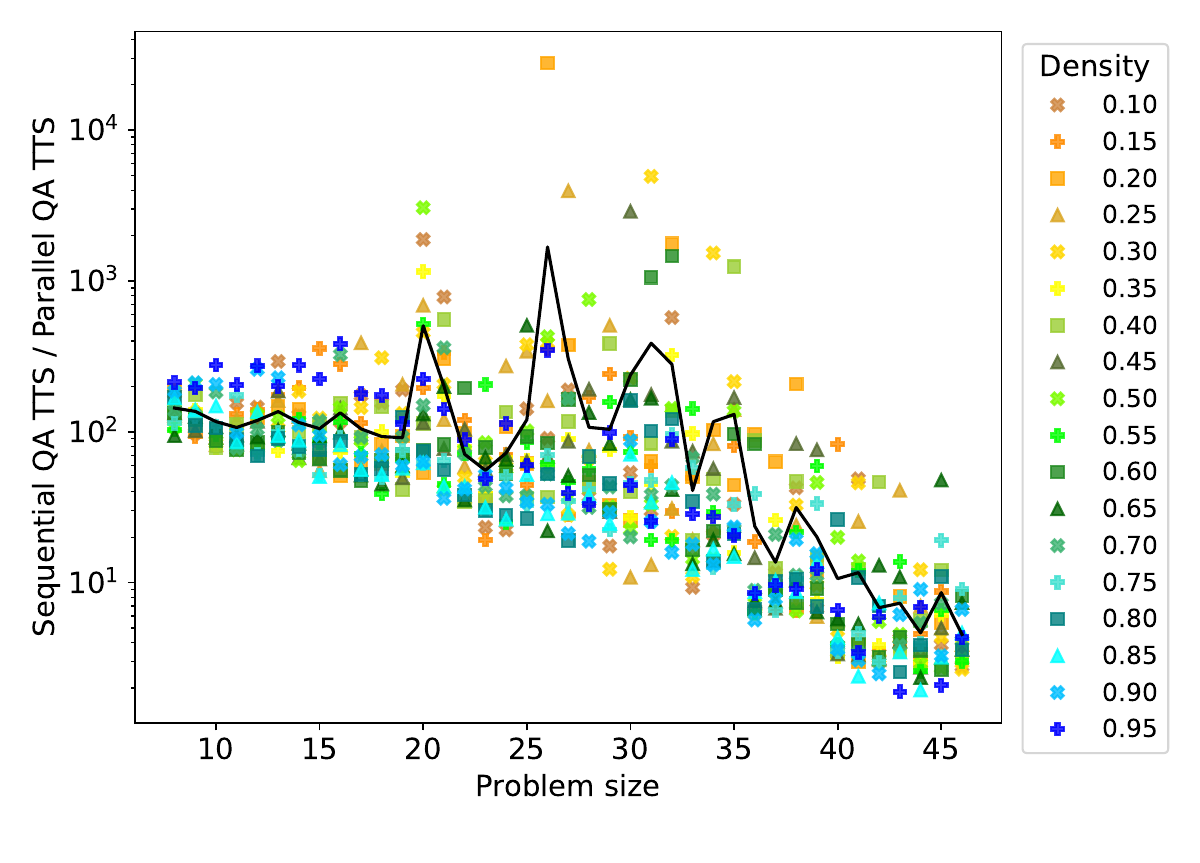}
    \put(46,-1){\textbf{(a)}}
    \end{overpic}\hfill
    \begin{overpic}[width=0.49\textwidth]{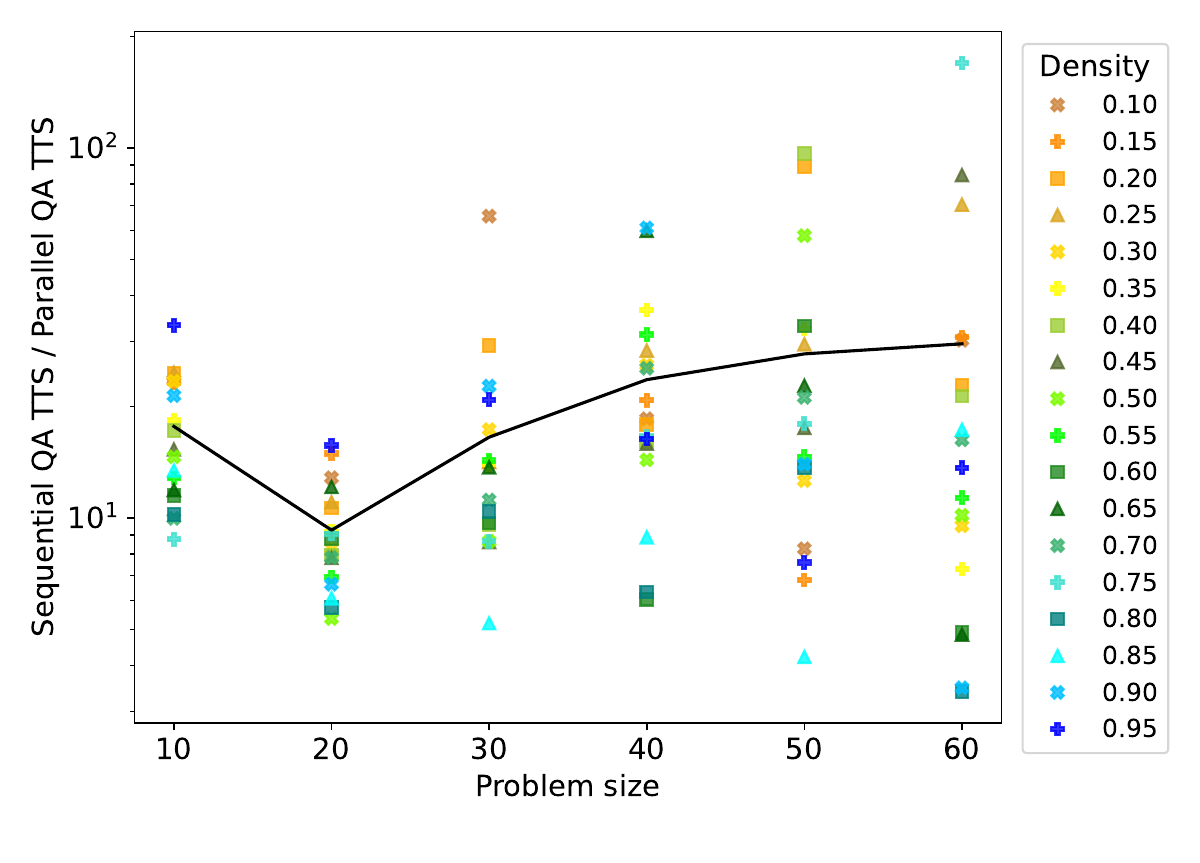}
    \put(46,-1){\textbf{(b)}}
    \end{overpic}
    \caption{TTS speedup of using parallel quantum annealing compared to sequential quantum annealing as a function of the problem size for D-Wave 2000Q (a) and D-Wave Advantage~1.1 (b). The TTS speedup is computed as the quotient of the sequential and parallel TTS metrics. Mean values are shown by the solid black line. Graph densities range from $0.10$ to $0.95$ (color coding shown in the legend).}
    \label{fig:TTS_comparison}
\end{figure}
Next, Figure~\ref{fig:TTS_comparison} compares the sequential and parallel quantum annealing efficiencies with respect to the TTS metric. The TTS speedup is computed as the quotient of the sequential and parallel TTS values. For DW~2000Q (a), we observe that the proposed parallel quantum annealing approach yields substantial speedups (an average of $152$-fold over all problem sizes considered) compared to solving the same problems sequentially on the D-Wave chip. A stratification by graph density is not observable in this case. As expected, the speedup is more pronounced for smaller problem sizes, as in this case more problems can be solved in parallel. For DW~Advantage, a considerable speedup (on average around $20$-fold) is observed over all problem sizes considered.

Note that the CPU unembedding time increases when unembedding larger datasets; meaning that for larger backends (and more variables used per anneal) the total unembedding time will increase. Therefore, the differences in TTS between the two backends seen in Figure~\ref{fig:TTS_comparison} are not only caused by the differences in GSP seen in Figure~\ref{fig:GSP_comparison}.

\subsection{Comparison with a fast classical solver}
\label{sec:classical}
\begin{figure}[t]
    \centering
    \begin{overpic}[width=0.49\textwidth]{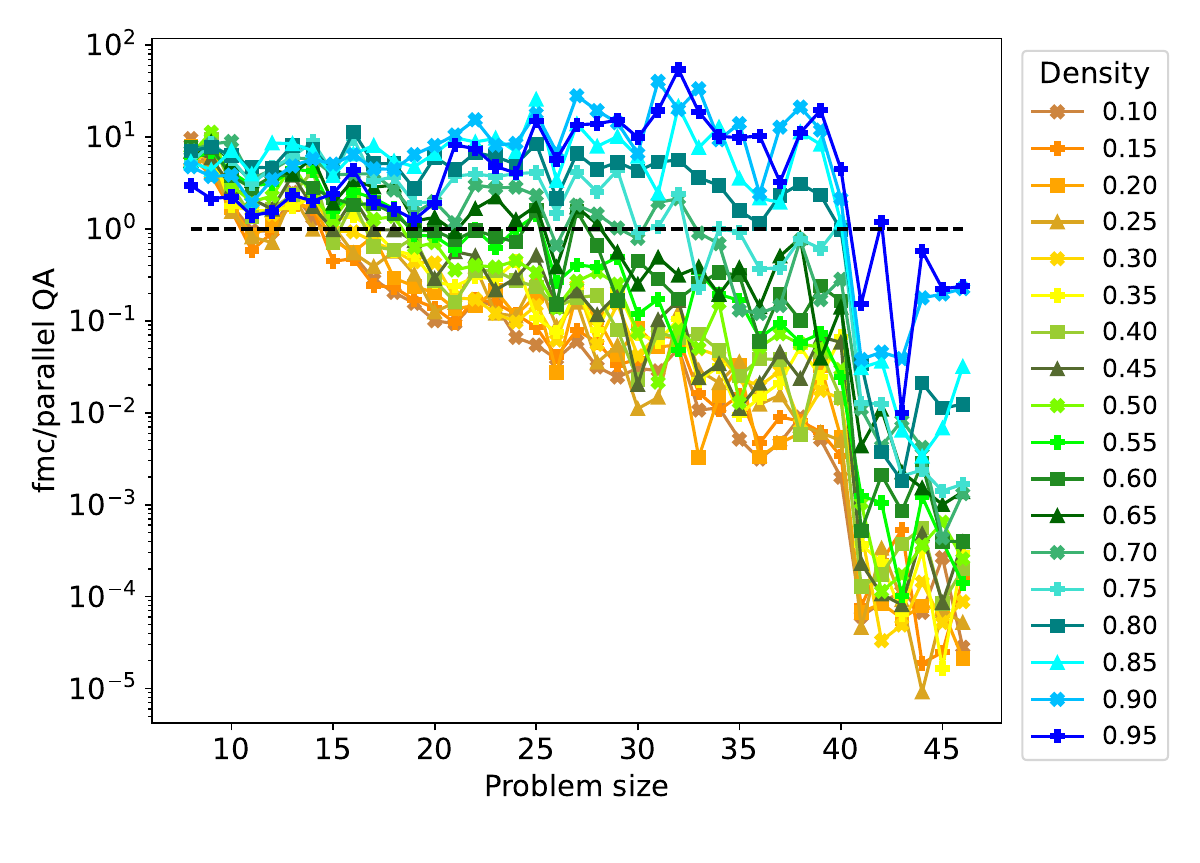}
    \put(46,-1.5){\textbf{(a)}}
    \end{overpic}\hfill
    \begin{overpic}[width=0.49\textwidth]{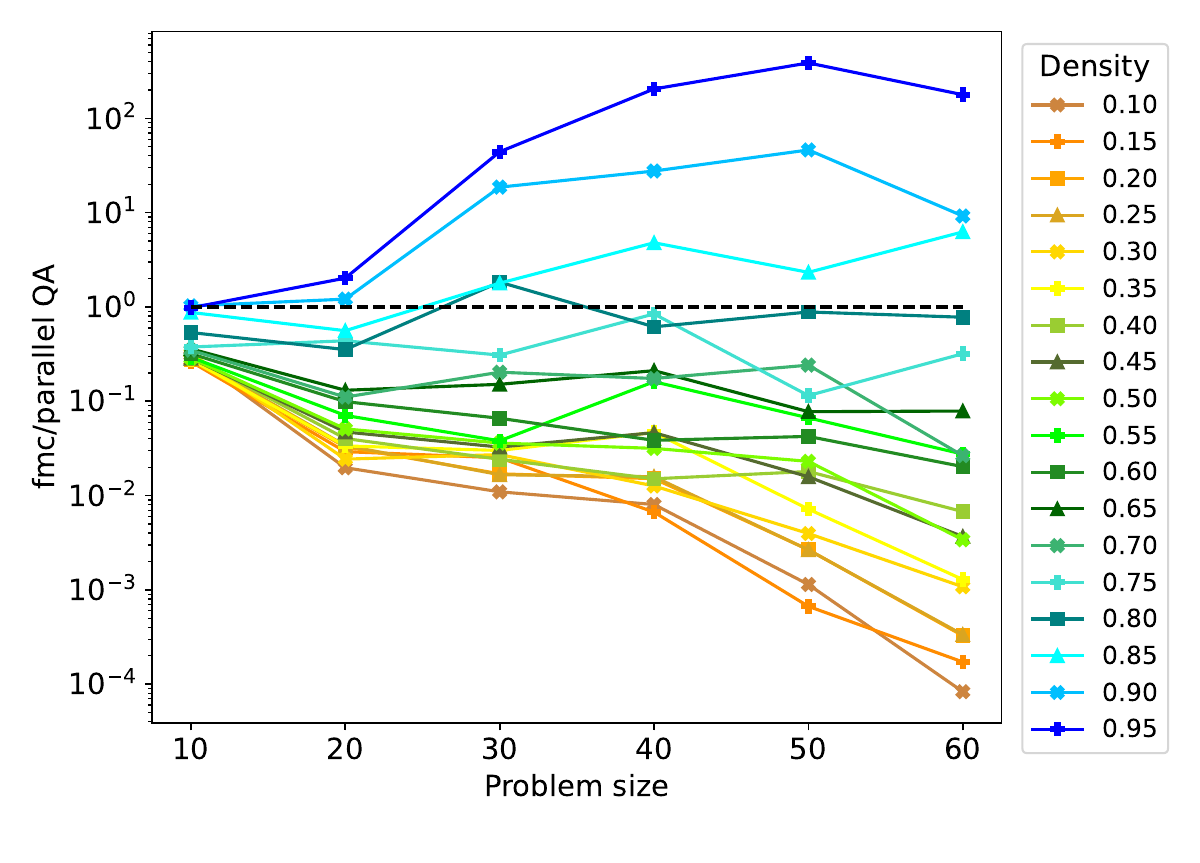}
    \put(46,-1.5){\textbf{(b)}}
    \end{overpic}
    \caption{Speedup of using parallel quantum annealing compared with the classical FMC solver as a function of the problem size for DW 2000Q (a) and DW Advantage (b). Log scale on the y-axis. Graph densities range from $0.10$ to $0.95$ (color coding shown in the legend).}
    \label{fig:FMC_comparison}
\end{figure}
We compare the performance of the proposed parallel quantum annealing approach with a classical solver for the Maximum Clique problem. We employ the \textit{fast maximum clique} (FMC) solver \cite{fmc}. FMC is run in exact mode throughout this experiment.

Figure~\ref{fig:FMC_comparison} shows the speedup of parallel quantum annealing over FMC. For the probabilistic parallel QA, we measure ensemble TTS (see Section~\ref{sec:tts}), whereas for the classical FMC solver, we measure CPU times (this is justified since TTS reduces to CPU time in the classical case). The speedup is computed as the quotient of the parallel TTS metric and the CPU process time used by FMC. For DW~2000Q (a), we observe that for both small problems (of any density) and for high densities up to problem size $40$, using parallel quantum annealing yields better TTS values than using FMC. In turn, FMC is superior to DW~2000Q for low density graphs. For problems exceeding size $40$, DW~2000Q is not the preferred choice (as seen in previous experiments), making FMC the faster choice for all densities. Interestingly, for DW~Advantage (b), we observe that D-Wave is not superior to FMC apart from very high densities. In turn, the DW~Advantage architecture does not suffer from a reduced accuracy for large problem sizes (compared to the DW~2000Q).

\subsection{Improving a single hard problem's TTS using parallel quantum annealing}
\label{sec:same_problem_improved_TTS}
Finally, we consider how parallel quantum annealing can be used to increase the GSP, and therefore TTS, of a single hard QUBO problem. Such problems need to be run sequentially many times on the quantum annealer to get even a single optimal solution. Our idea is to run multiple replicas (identical copies of the QUBO) simultaneously using parallel quantum annealing.

\begin{figure*}
    \centering
    \begin{overpic}[width=0.49\textwidth]{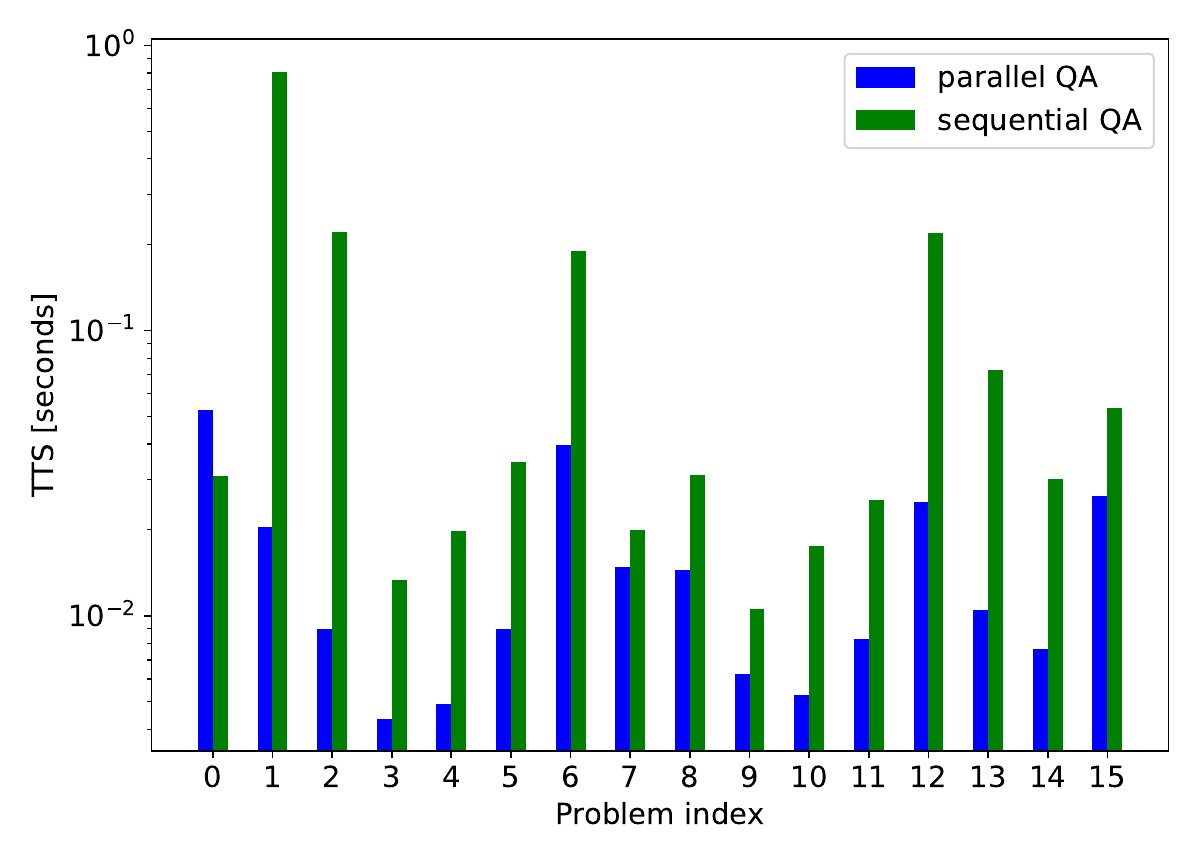}
    \put(53,-1.5){\textbf{(a)}}
    \end{overpic}\hfill
    \begin{overpic}[width=0.49\textwidth]{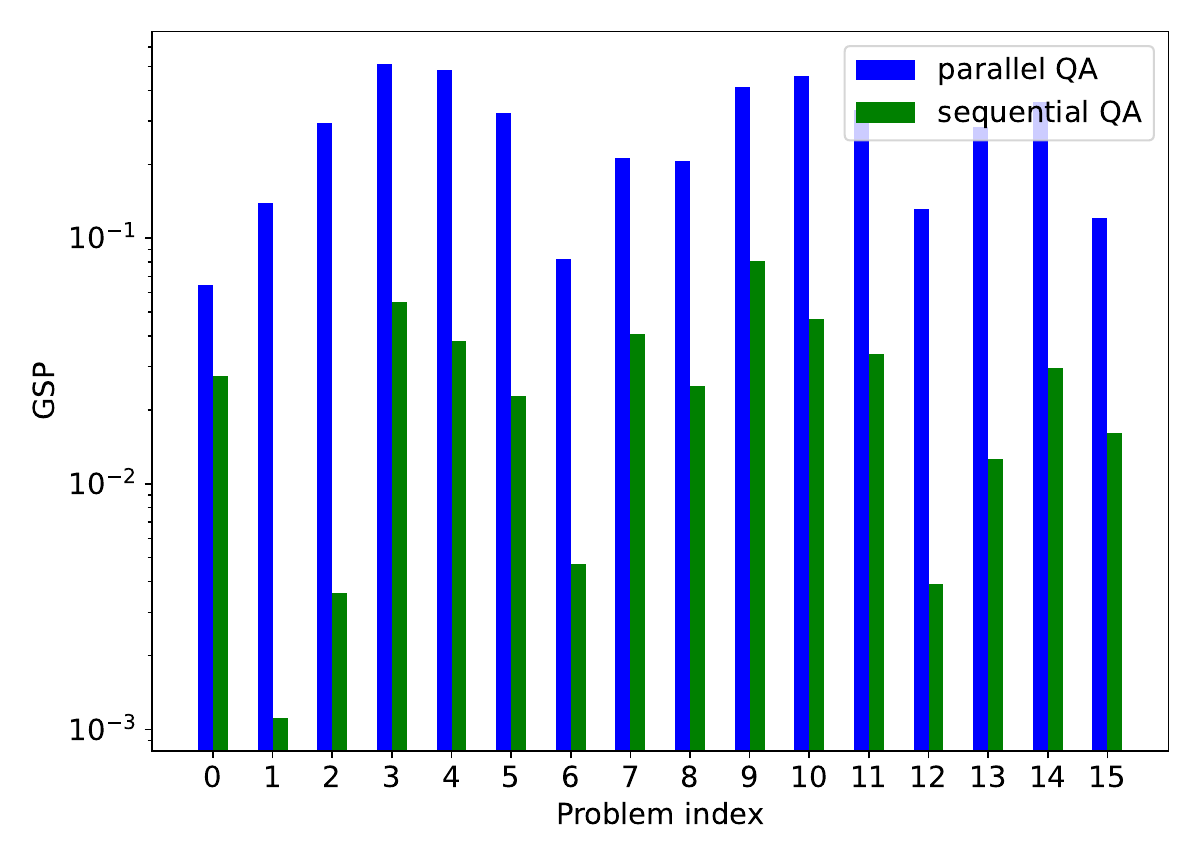}
    \put(53,-1.5){\textbf{(b)}}
    \end{overpic}
    \caption{TTS (a) and GSP (b) for the $15$ test problems we consider, both for parallel quantum annealing (blue) and sequential quantum annealing (green). Log scale on the y-axis.}
    \label{fig:same_problem_TTS_and_GSP}
\end{figure*}

We compare sequential quantum annealing and parallel quantum annealing on DW~Advantage for a fixed set of $16$ random graphs with $N=40$ nodes and $p=0.5$. Each problem is first solved sequentially, and afterwards solved in parallel by embedding it $16$ times onto the hardware. Results are shown in Figure~\ref{fig:same_problem_TTS_and_GSP}, where we see that parallel quantum annealing can reduce TTS due to improved GSP. Over all problems considered, the average TTS speedup (the quotient of sequential QA TTS and parallel QA TTS) is $7$, and the average GSP increase (the quotient of parallel QA GSP and sequential QA GSP) is $10$. Note that problem $0$ had worse TTS in the parallel quantum annealing case. This is due to an increased cost in the CPU unembedding time in the parallel case. 

To compute the TTS measure of the parallel implementation, we use eq.~\eqref{eq:tts1}, and not eq.~\eqref{eq:tts2}. This is due to the fact that there are not $K$ distinct problem being solved in parallel; instead, we are solving the same problem $K$ times during the same annealing cycle. Therefore, $p$ is simply the proportion of anneals that found the optimal solution. Note, however, that we can have cases where the ground state solution is found multiple times in the same anneal (we do not distinguish between finding the ground state once or more than once in the embedded problems).

\section{Discussion}
\label{sec:discussion}
This article investigates a proposed method called parallel quantum annealing, which allows one to solve many independent problems on a quantum annealer during the same annealing cycle. The idea is justified by the fact that independent QUBOs, that is those not sharing any variable, can be added together to yield a combined QUBO that preserves the bitstring solution of each, and has a new ground state probability that is the sum of the individual ground state probabilities.

We evaluate the proposed approach with respect to both accuracy and the TTS metric on both DW 2000Q and DW Advantage. We compare to sequential quantum annealing (i.e., solving the same problems sequentially on D-Wave), and the classical FMC solver. We demonstrate that while the solution accuracy is slightly decreased for simultaneously solved problems, parallel quantum annealing can yield a considerable speedup of up to two orders of magnitude. Interestingly, using the newer DW~Advantage does not yield more pronounced speedups in TTS than the previous generation DW~2000Q.

Notably, parallel quantum annealing is different from classical parallelism in that parallel quantum annealing solves multiple problems on a single quantum chip, whereas classical parallel computing solves independent problems at the same time, but using different physical circuits. Therefore, for a sufficiently large quantum annealer (and assuming the solution quality does not suffer), a very large number of problems \emph{could} be solved in a very short amount of time (determined by the annealing time and the number of anneals). 

The number of qubits in new quantum annealers have been steadily increasing, but there has been a concern that higher error rates may prevent such machines from solving very large problems. Parallel quantum annealing may be an alternative to making future larger machines usable.

This work leaves scope for future research avenues:
1. This work only considers solving Maximum Clique problems simultaneously on D-Wave. Investigating other important NP-hard problems remains for future work. In particular, the QUBOs being solved in parallel do not have to stem from the same problems, and they do not need to be of equal size.
\\2. Optimizing the annealing parameters (such as annealing time, chain strengths, etc.) for DW Advantage might significantly improve its performance.
\\3. It remains to investigate why parallel quantum annealing unfairly finds the ground states of some problems at higher rates than others; and how this effect can be mitigated. A related question is how fairly parallel quantum annealing samples the degenerate ground states of the embedded problems.

\section{Methods}
\label{sec:methods}
This section discusses some of the elements of the proposed parallel quantum annealing technique, in particular the QUBO formulation of the Maximum Clique problem (Section~\ref{sec:maxcliquequbo}), combining multiple small QUBOs into a single one (Section~\ref{sec:qubos}), the choice of the embedding (Section~\ref{sec:embedding}), and the specifics of the D-Wave execution (Section~\ref{sec:dwave_procedure}).

\subsection{Maximum Clique QUBO}
\label{sec:maxcliquequbo}
The Maximum Clique QUBO we use is given by~\cite{Chapuis2019, Lucas2014}
\begin{align}
    H_{MC} = -A\sum_{v \in V} x_v + B\sum_{(u,v) \in \overline{E}} x_u x_v,
    \label{eq:MC}
\end{align}
where the constants can be chosen as $A=1$, $B=2$ (see~\cite{Lucas2014}).

\subsection{Combining QUBOs of independent problems}
\label{sec:qubos}
Consider two QUBOs $H_1(x_1,\ldots,x_n)$ and $H_2(x_{n+1},\ldots,x_m)$ with respective minimum solutions $x_1^\ast,\ldots,x_n^\ast$ and $x_{n+1}^\ast,\ldots,x_m^\ast$. Since both QUBOs do not share any variable, the ground state of $H_1+H_2$, which is now a function in $x_1,\ldots,x_n,x_{n+1},\ldots,x_m$, is precisely the sum of the ground states of $H_1$ and $H_2$, and the minimum bitstring yielding the ground state will be $x_1^\ast,\ldots,x_n^\ast,x_{n+1}^\ast,\ldots,x_m^\ast$. This idea lays at the heart of parallel quantum annealing.
Given we have an embedding that allows us to place $H_1$ and $H_2$ simultaneously on the quantum chip, we can therefore embed and solve $H_1$ and $H_2$ in one D-Wave call. Naturally, $H_1$ and $H_2$ do not need to be QUBOs of the same type of optimization problem, and neither do they need to be of equal size. Generalization to a larger number of QUBOs is straightforward.

The D-Wave hardware has a hardware precision limit \cite{Pudenz2015, dorband2018extending, ice_precision_limit} for the provided problem coefficients. Therefore, when constructing these independent QUBOs, another point to consider is how the minimum and maximum range of these independent QUBOs compare to each other. If one of these QUBOs' coefficients dominate all of the other QUBOs, then most of the QUBO coefficients will effectively be lost in noise. Therefore, it might be necessary to normalize the minimum and maximum QUBO coefficients across all of the QUBOs that are being solved at the same time. In the experiments shown in Section~\ref{sec:experiments}, all of the problems are Maximum Clique QUBOs, which always have the same minimum and maximum coefficient range - therefore no coefficient normalization is used.

\subsection{Choice of the embedding}
\label{sec:embedding}
We aim to create multiple disjoint minor-embeddings of a complete graph of size $N$ onto the DW~2000Q and the DW~Advantage hardware. This allows us to embed arbitrary QUBOs of size $N$ onto the hardware connectivity graph, meaning that we can solve multiple problems of arbitrary structure (up to size $N$) in parallel in a single backend call.

\begin{figure*}
    \centering
    \begin{overpic}[width=0.49\textwidth]{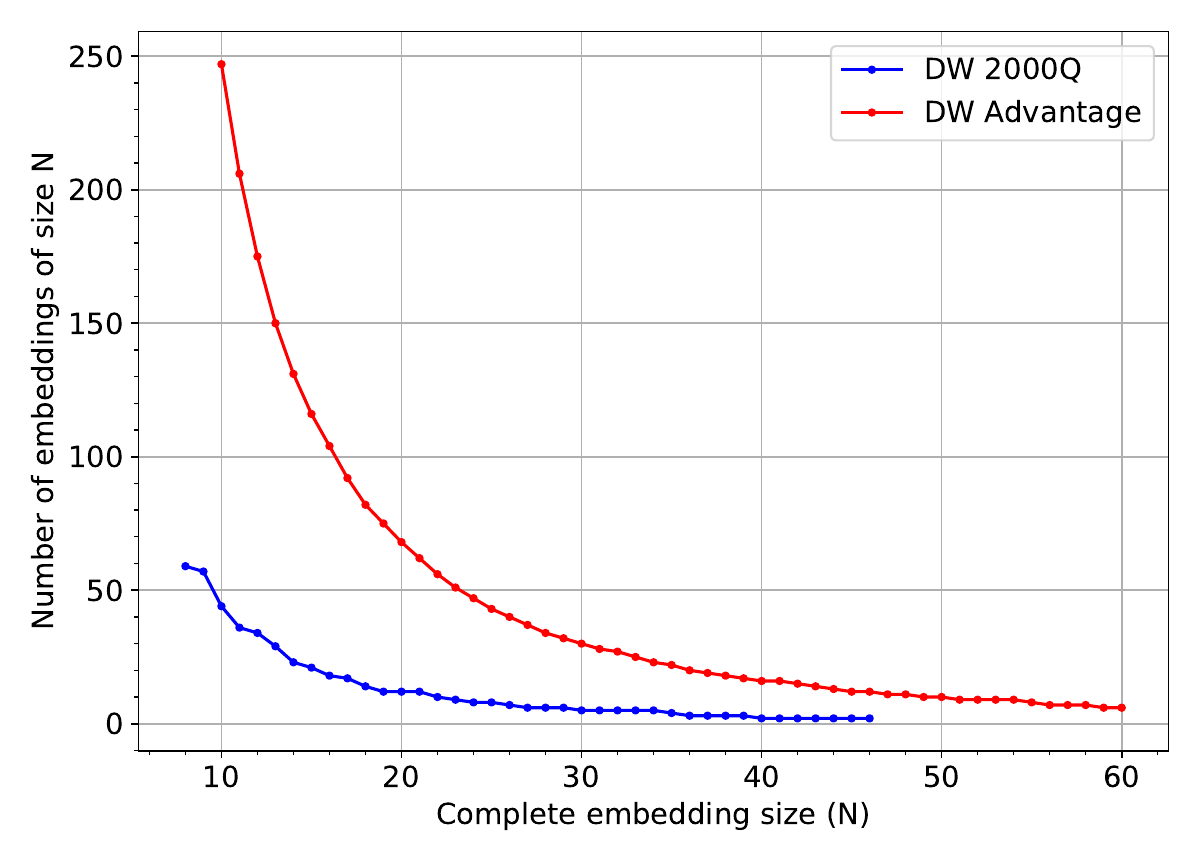}
    \put(53,-1.5){\textbf{(a)}}
    \end{overpic}\hfill
    \begin{overpic}[width=0.49\textwidth]{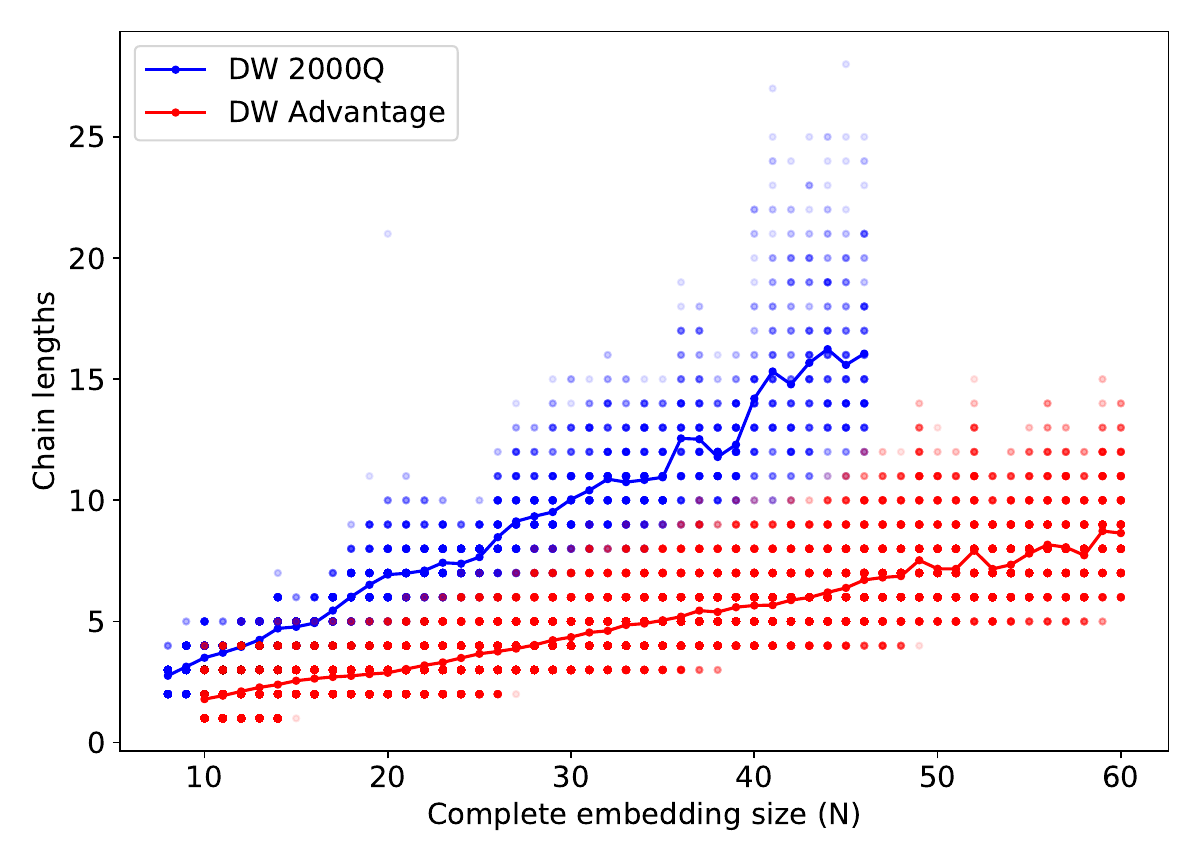}
    \put(53,-1.5){\textbf{(b)}}
    \end{overpic}
    \caption{(a): Number of embeddings as a function of the embedding size (clique size) $N$ for DW~2000Q (blue) and DW~Advantage (red). (b): Scatter plot of the chain lengths occurring in the embeddings as a function of the embedding size $N$, with the average depicted as a line. For DW~2000Q, we consider cliques of sizes $N \in \{8,\ldots,46\}$, while $N \in \{10,\ldots,60\}$ for DW~Advantage.}
    \label{fig:embedding_sizes}
\end{figure*}

We compute separate embeddings for clique sizes $N$ in the range of $N \in \{8,\ldots,46\}$ for DW~2000Q, and $N \in \{10,\ldots,60\}$ for DW~Advantage. As shown in Figure~\ref{fig:cliques}, ideally the embeddings should be chosen such that both the number of cliques (complete graphs) embedded on the quantum hardware, as well as the physical distance between the embeddings is maximized. The latter is conjectured to decrease spurious interactions between qubits of different QUBOs on the chip, thus helping to increase the solution quality.

The disjoint embeddings for both DW~2000Q and DW~Advantage are computed using the method \textit{minorminer}~\cite{minorminer, Cai2014practical}. In particular, we supply the graph to be embedded to minorminer as simply the union of the disjoint cliques we wish to embed. This is done for all clique sizes $N$ for which an embedding can be found by minorminer. Figure~\ref{fig:embedding_sizes} shows both the number of embeddings we are able to place on the D-Wave quantum chip, as well as the chain lengths occurring in those embeddings for both DW~2000Q and DW~Advantage.

Supplying minorminer with a graph consisting of the disjoint cliques we wish to embed is sub-optimal. In particular, when computing the embeddings, we would ideally be optimizing for the separation between the embeddings on the hardware. This remains for future work, thus close physical proximity of our embeddings may cause leakage between qubits.

As described in Section~\ref{sec:intro}, after annealing, we need to unembed all chains. For this, we tried both the \textit{majority vote} method and the \textit{random weighted} method. All results shown in this paper use the \textit{random weighted} unembedding method.

\subsection{D-Wave procedure}
\label{sec:dwave_procedure}
For each of the number of disjoint embeddings we fixed in Section~\ref{sec:embedding}, we generate an Erd\H{o}s–R\'{e}nyi random graph $G(n,p)$ of size $n=N$, varying edge probability $p$ from $0.10$ to $0.95$ in increments of $0.05$ such that each random graph is connected, does not have any degree $0$ nodes, and is not a full clique itself. Then, we compute the exact Maximum Clique solution using FMC~\cite{fmc} and Networkx~\cite{networkxCliques, Bron1973, Tomita2006, Cazals2008} (the Networkx solver is used to find any degenerate Maximum Clique solutions). This step is necessary in order to find the CPU time used by FMC, as well as determine the exact solution(s) for the purpose of computing ground state probability (GSP), which is used to compute the quantum annealer Time-to-Solution (TTS) metric.

Next, we arbitrarily assign each of the random graphs to one of the disjoint embeddings. Then, the Maximum Clique QUBO is computed for each of these random graphs, and each QUBO is then embedded onto its assigned embedding. For example, Figure~\ref{fig:cliques} (a) shows that on DW~2000Q for $N=20$, we generate $12$ random graphs of size $20$ and then embed those 12 Maximum Clique QUBOs onto each of their respective (independent) embeddings. Then, this is repeated for each density.

Once the problems are embedded, we call the D-Wave backend using the parameter setting of Section~\ref{sec:tuning}. In order to get reliable results for the TTS computation, we call the backend $100$ times for each problem, and each backend call requests $1,000$ anneals. This results in $100,000$ samples in total per random graph. From these results, we first unembed the samples using the \textit{random weighted} method, and then we  compute how many of the $100,000$ samples correctly found the Maximum Clique solution. Using this information, we then compute the \textit{ensemble TTS} metric using eq.~\eqref{eq:tts2}.

Next, we apply the sequential quantum annealing method; meaning that we solve each of those QUBOs using \emph{separate} backend calls. Importantly, each QUBO still uses the same physical hardware embedding as it did in the parallel case, so that differences in the embeddings will not change solution quality. In the example of Figure~\ref{fig:cliques} (a), we would perform the $100$ backend calls for each one of the $12$ distinct problems, without the other $11$ embedded into the chip. This procedure is then repeated for all graph densities as well. Once again, we unembed using the \textit{random weighted} method, and we also compute GSP for each of the problems. This procedure allows us to directly compare the solution quality for each of these problems when solving them separately and in parallel. Using this data, we can compute the \textit{TTS} using eq.~\ref{eq:tts1} for each of the random graphs. Thus, the \emph{total TTS} for a group of graphs that were solved sequentially is the sum of each of their individual TTS values (in contrast to using parallel quantum annealing where we compute the \textit{ensemble TTS} using eq.~\ref{eq:tts2}).

It is important to note that, especially for larger clique sizes (e.g., cliques of size $40$ and larger for DW~2000Q, and clique sizes $50$ onward for DW~Advantage), not all problems can be solved to optimality by the quantum annealer. In this case, we encounter missing information in the computation of the TTS metric. We mitigate this problem by attempting to solve another randomly generated problem instead until the TTS metric can be computed. However, in settings where particular graphs need to be solved (as opposed to the setting with random graphs we consider), it is likely that D-Wave fails for large problems solved using a fixed embedding~\cite{Barbosa2021prediction}. These failures can be attributed to longer chain lengths leading to broken chains, and therefore worse solution quality. Our results in section~\ref{sec:same_problem_improved_TTS} show a possible solution to this problem; that is using many different embeddings to solve the same problem in parallel might allow D-Wave to more consistently find optimal solutions for large problem sizes.

\section*{Acknowledgments}
\label{sec:Acknowledgments}
The research presented in this article was supported by the Laboratory Directed Research and Development program of Los Alamos National Laboratory under the project number 20190065DR. The work of Hristo Djidjev has been also partially supported by Grant No.~BG05M2OP001-1.001-0003, financed by the Science and Education for Smart Growth Operational Program (2014-2020) and co-financed by the European Union through the European Structural and Investment Funds.

\section*{Code and data availability}
\label{sec:code}
The code and the data (e.g., the embeddings) are available at
\begin{center}
    \url{https://github.com/lanl/Parallel-Quantum-Annealing} \cite{parallel_QA_software}
\end{center}

\printbibliography

\end{document}